\def   \ni {\noindent}

\def   \ssk {\vskip  5truept}

\def   \bsk {\vskip 15truept}

\def   \newline {\hfil\break}

\newcommand{\rmn}[1] {\mathrm{#1}}

\documentstyle[epsfig]{article}
\begin{document}

%
\def\la{\mathrel{\mathchoice {\vcenter{\offinterlineskip\halign{\hfil
$\displaystyle##$\hfil\cr<\cr\sim\cr}}}
{\vcenter{\offinterlineskip\halign{\hfil$\textstyle##$\hfil\cr
<\cr\sim\cr}}}
{\vcenter{\offinterlineskip\halign{\hfil$\scriptstyle##$\hfil\cr
<\cr\sim\cr}}}
{\vcenter{\offinterlineskip\halign{\hfil$\scriptscriptstyle##$\hfil\cr
<\cr\sim\cr}}}}}
\def\ga{\mathrel{\mathchoice {\vcenter{\offinterlineskip\halign{\hfil
$\displaystyle##$\hfil\cr>\cr\sim\cr}}}
{\vcenter{\offinterlineskip\halign{\hfil$\textstyle##$\hfil\cr
>\cr\sim\cr}}}
{\vcenter{\offinterlineskip\halign{\hfil$\scriptstyle##$\hfil\cr
>\cr\sim\cr}}}
{\vcenter{\offinterlineskip\halign{\hfil$\scriptscriptstyle##$\hfil\cr
>\cr\sim\cr}}}}}
\def\degr{\hbox{$^\circ$}}
\def\arcmin{\hbox{$^\prime$}}
\def\arcsec{\hbox{$^{\prime\prime}$}}

\hsize 5truein
\vsize 8truein
\font\abstract=cmr8
\font\keywords=cmr8
\font\caption=cmr8
\font\references=cmr8
\font\text=cmr10
\font\affiliation=cmssi10
\font\author=cmss10
\font\mc=cmss8
\font\title=cmssbx10 scaled\magstep2
\font\alcit=cmti7 scaled\magstephalf
\font\alcin=cmr6 
\font\ita=cmti8
\font\mma=cmr8
\def\ref{\par\noindent\hangindent 15pt}
\null


\title{\ni 
Tests of Gaussianity of CMB maps}                                              

\bsk \bsk
\author{\ni Enrique Mart\'\i nez-Gonz\'alez$^1$, R. 
Bel\'en Barreiro$^{1,2}$, Jos\'e M. Diego$^{1,2}$, Jos\'e Luis Sanz$^1$, 
Laura Cay\'on$^1$, Joseph Silk$^{3}$, Silvia Mollerach$^4$ \& Vicent J. 
Mart\'\i nez$^5$}                 
                                      
\bsk
\affiliation{$^1$ Instituto de F\'\i sica de Cantabria, Consejo Superior de 
Investigaciones Cient\'\i ficas-Universidad de Cantabria, Avda.
Los Castros s/n, 39005 Santander, Spain\\
\indent $^2$ Departamento de F\'\i sica Moderna, Universidad de Cantabria, 
Avda. Los Castros s/n, 39005 Santander, Spain\\
\indent $^3$ Center for Particle Astrophysics, Department of Astronomy and 
Physics, University of California, Berkeley CA 94720-7304, USA\\
\indent $^4$ Departamento de F\'\i sica, Universidad Nacional de La Plata,
cc 67, La Plata 1900, Argentina\\
\indent $^5$ Departamento de Astronom\'\i a y Astrof\'\i sica, Universidad de
Valencia, E-46100 Burjasot, Valencia, Spain 
}                                                
\bsk
\baselineskip = 12pt

\abstract{ABSTRACT \ni
We study two different methods to test Gaussianity in CMB maps. One
of them is based on the partition function and the other on the morphology of 
hot and cold spots.
The partition function contains information on all the 
moments and scales, being a useful quantity to compress the large data sets 
expected from future space missions like Planck. In particular, it contains 
much richer information than the one available through the radiation power 
spectrum. The second method utilizes morphological properties
of hot and cold spots such as the eccentricity and number of spots
in CMB maps. We study the performance of both methods in detecting  
non-Gaussian features in small scale CMB simulated
maps as those which will be provided by the Planck mission.

}                                                    
\bsk
\baselineskip = 12pt

\bsk
\baselineskip = 12pt


\text{\ni 1. INTRODUCTION
\ssk
\ni
Future CMB experiments like the Planck mission will provide very large data 
sets at small angular scales with a high sensitivity. Their
analysis will require the development of new and sophisticated  methods
capable of managing the data, performing an optimal separation of the
different components (CMB,
Galactic and extragalactic foregrounds),  extracting their statistical
properties and finally determining the fundamental cosmological
parameters. Two methods for foreground removal have already been
proposed based on 
Maximum Entropy (Hobson et al. 1998, 1999a) and Wiener filter (Tegmark and
Efstathiou 1996). The power spectrum of the temperature fluctuations
is the most widely used quantity in order to compress the data and 
obtain the cosmological parameters. This quantity only carries
information on the second moment of the distribution and completely 
characterizes the statistical properties of the data if these are
Gaussian distributed. However, at present there
\vfill\eject
\noindent is no experimental evidence for that statistical behaviour in the
small scale temperature fluctuations. Moreover, topological defect
models predict non-Gaussian temperature fluctuations. It is therefore
very important to explore other quantities that can be sensitive to
non-Gaussian features and can test the assumption of
Gaussianity usually made for the cosmological signal.

Several studies of the Gaussian character of the CMB
signal based on different statistical and morphological quantities
have already been performed.
They include extrema correlation function (Kogut et al. 1996,
Barreiro et al. 1998), number of hot and cold spots (Coles and Barrow 1987),
genus (Gott et at. 1990, Kogut et al. 1996), 
bispectrum (Ferreira et al. 1998, Heavens 1998), 
wavelets (Hobson et al. 1999b, 
Pando et al. 1999) and multifractals (Pompilio et
al. 1995). 

In this paper we concentrate on the recently proposed partition
function of the CMB temperature and also on the morphology of the
temperature extrema. 
The usefulness of the partition function to study the
Gaussianity of the CMB has already been pointed out in a previous work
(Diego et al. 1999). The study of morphological properties of CMB extrema, such
as number, curvature and eccentricity, has been done by Barreiro et al. (1997)
for Gaussian temperature fluctuations. Here we will probe the
practical performance of both methods by testing their abilility to find the
non-Gaussian features of very weakly perturbed Gaussian random fields.
In particular we will simulate small scale temperature maps
by slightly perturbing Gaussian ones 
and generating a small: a) skewness or b) kurtosis. 

\bsk
\ni 2. TESTS OF GAUSSIANITY 
\ssk
\ni 

\ni
2.1 The partition function
\ssk
\ni

The partition function has recently been introduced to study CMB data
by Diego et al. (1999). Below we concentrate on the application of this 
function to the study of Gaussianity.\\
The partition function is defined as follows:
\begin{equation}
 Z(q,\delta) =\sum_{i=1}^{N_{\rm boxes}(\delta)} \mu_i(\delta)^q
\end{equation}
The quantity
$\mu_i(\delta)$ is called the {\it measure}, it is a function
of $\delta$ which is the size or scale of the boxes used to cover the 
sample. The boxes are labeled by $i$ and  $N_{\rm boxes}(\delta)$ 
is the number of
boxes (or cells) needed to cover the map when the grid with
resolution $\delta$ is used. The exponent $q$ is a continuous 
real parameter that plays the role of the order of the moment of
the measure.\\ 
Let us consider a CMB map of $N$ pixels. Now the map is divided
in boxes of size $\delta\times\delta$ pixels and the measure
$\mu_i(\delta)$ is computed in each one of the resulting boxes.
Changing both, $q$ and $\delta$, one calculates the function
$Z(q,\delta)$. We would like to emphasize that the calculation of 
$Z(q,\delta)$ is $O(N)$.\\ 
One is free to make any choice of the measure $\mu(\delta)$ 
provided that  several conditions are satisfied, 
the most restrictive  being $\mu_i(\delta) \ge 0$.  
There are no general rules to decide
which is the best choice. To test Gaussianity we use the following 
measure:
\begin{equation} 
\mu_i(\delta)=\frac{1}{T_*} \sum_{{\rm pix}_j \in {\rm box}_i}
 T_{{\rm pix}_j}.
\end{equation} 
Thus the measure in the box $i$ is the sum of the {\it absolute}
temperatures $T_{\rm pix}$ of the pixels  
inside the box in units of Kelvin. 
The constant $T_*$ is a normalization constant. 
The measures are interpreted as probabilities and they have to be 
normalized, i.e $\sum_i \mu_i = 1$. So $T_*$ is simply the sum of 
the absolute temperatures over all pixels and therefore is a constant 
for all boxes and scales.\\

Summarizing, $Z(q,\delta)$ contains information at different scales
and moments. The multi-scale information gives an idea of the
correlations in the map, meanwhile the moments are sensitive to
possible asymmetries in the data, as well as some deviations from
Gaussianity. In order to study the last property let us introduce the
generating function:
\begin{equation}
G_{\mu}(q)=\langle e^{q\mu} \rangle=\frac{1}{N_{\rmn boxes}(\delta)}
\sum_{i=1}^{N_{\rmn boxes}(\delta)} e^{q\mu_i(\delta)}.
\end{equation}
Defining a new function $H(q)$ in the following way,
\begin{equation}
H(q)=lnG_{\mu}(q)-q\langle \mu\rangle - \frac {q^2\sigma^2_{\mu}}{2},
\end{equation}
this function vanishes for any value of $q$ when a Gaussian 
distributed variable is considered. In the case of the CMB temperature, if the
universe went through a phase of inflation then the fluctuations generated in
the CMB will be Gaussian. However, due to the cosmic and sample variance 
$H(q)$ for a given realization will not 
exactly vanishes, as we will see in the next section.  

\begin{figure}
\centerline{\psfig{file=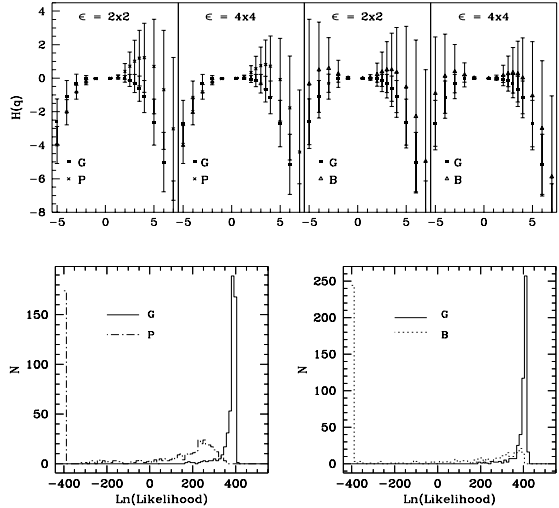, height=7.5cm, width=8cm}}
\caption{FIGURE 1. Top: The function $H(q)$ for two different sizes of the 
boxes. Bottom: Distribution  of $Ln(Likelihood)$ corresponding to $H(q)$ for 
the Gaussian and non-Gaussian models (see text).
}
\end{figure}


\ssk
\ni
2.2 The morphology of hot and cold spots
\ssk
\ni

The properties of the CMB temperature peaks for the Gaussian case have been 
recently studied in Barreiro et al. (1997). The local description of maxima 
involves the 
second derivatives of the field along the two principal directions. As usual,
the curvature radii are defined by $R_1=[-\Delta^{''}_1(max)/2]^{-1/2}$ and
$R_2=[-\Delta^{''}_2(max)/2]^{-1/2}$, where $\Delta$ is the temperature field 
normalized to the rms fluctuations. Then, we can associate two invariant 
quantities to any maximum: the Gaussian curvature $\kappa$ and the eccentricity
$\epsilon $ given by

\begin{equation}
\kappa = \frac {1}{R_1R_2},\ \ \ \epsilon = \left[1 -
\left(\frac{R_2}{R_1}\right)^2\right]^{1/2}.
\end{equation}

In addition to the Gaussian curvature and the eccentricity another interesting
property is the number of maxima (minima) $N_{>\nu}$ above (below) a given 
temperature threshold $\nu$ ($-\nu$) in CMB maps. For symmetric distributions, 
such as the 
Gaussian one, the number of maxima and minima are statistically 
undistinguishable. Thus, any asymmetry present would be a mark of 
non-Gaussianity. Instead of using those 
morphological quantities for extrema in practice it is simpler to use 
excursion sets, i.e. hot and
cold spots regions above and below a given temperature threshold. Therefore, 
we will count the number of excursion sets from the maps. To estimate the 
eccentricity of the excursion sets we fit each region (formed by at least 
6 pixels) with an ellipse of equal area as the corresponding excursion set.
Finally, the Gaussian curvature is a more difficult quantity to estimate 
from the simulated maps and we will 
leave it for a forthcoming paper.
\bsk
\ni
3. SIMULATIONS
\ssk
\ni

We perform small scale simulations of size $12.8\times 12.8$ degrees with 
pixel and antenna sizes similar to 
those of the best Planck channels, i.e. $1'.5$ and $5'$ respectively. The 
amount of instrumental noise in the maps provided by this mission
is expected to be much smaller than the CMB signal, then, as a first 
approximation, it is not considered in this work. We simulate maps with 
Gaussian 
and non-Gaussian statistical properties, that will be used to test the 
previously discussed 
methods. The properties of the Gaussian 
perturbations are completely defined by the standard inflationary model. 
Although many models have been proposed for which non-Gaussian temperature 
perturbations are produced,  
however all their statistical properties are either not given or are very 
complicated to simulate. We then consider ``generic'' non-Gaussian 
perturbations derived from Gaussian ones using transformations given by 
Cole and Weinberg (1992).
The non-Gaussian temperature fields are 
derived from Gaussian fields in two ways. First, we perform the 
following 
transformation to an uncorrelated 2D Gaussian field G: $exp(aG)-exp(a^2/2)$ 
with $a$ a
parameter which recovers the Gaussian case when $a\rightarrow 0$. 
An uncorrelated 
non-Gaussian field with zero mean is thus obtained. 
The radiation 
power spectrum corresponding 
to a standard CDM model normalized to COBE/DMR is introduced by rescaling the 
amplitude of each Fourier mode to that spectrum. By rescaling the power 
spectrum the 1-point distribution is modified with respect to the uncorrelated
one, but it is still non-Gaussian. Second, we consider the 
transformation: $G[exp(aG)+exp(-aG)]$. Then the same radiation power spectrum
is introduced as before. Finally the maps are smoothed with a FWHM$=5'$. 
In the first case we construct a non-Gaussian field 
whose distribution is skewed with a positive tail (P) and in the second one the
distribution is broadened with both positive and negative tails (B). For
both non-Gaussian fields we consider small enough values of the parameter 
$a$ such that their 1-point distributions are undistinguishable from a 
Gaussian. In addition their power spectra are undistinguishable from the
one of the Gaussian CDM model by construction. 
\bsk
\ni
4. RESULTS
\ssk
\ni

The results of comparing two sets of simulated maps, one corresponding to 
Gaussian fluctuations and the other to one of the previous non-Gaussian 
fields, using the methods described in section 2 are summarized in figures 1 
and 2 for the functions
$H(q)$ and $N_{>\nu}$, respectively. We have considered a value $a=0.8$ for 
which it is impossible to distinguish between the corresponding 1-pdf. 
For the set of values of $H(q)$ ($N_{>\nu}$) given by the top of figure 1 (2)
we have constructed
a likelihood function whose values for each of the simulated Gaussian and 
non-Gaussian maps (compared to the Gaussian model) are shown in the bottom
of figure 1 (2). We see that a positive result, i.e. it is posible to 
distinguish between the two distributions for most of the simulations,
is obtained for the P non-gaussian field for both the $H(q)$ and $N_{>\nu}$
quantities. In the case of the B non-gaussian field the difference is less 
pronounced but it is still possible to discriminate between both models in a 
number of cases.\\ 
We have also performed a similar test for
the eccentricity $\epsilon_{>\nu}$ but in this case it is not possible to
discriminate the Gaussian and any of the non-Gaussian fields.

Summarizing, the function $H(q)$ as well as $N_{>\nu}$ are sensitive 
quantities with which to probe the possible non-Gaussian features that can be
present in future high sensitivity maps. On the contrary, the eccentricity is
not a good discriminator of Gaussianity at least for the small maps 
considered in the present work.  

\begin{figure}
\centerline{\psfig{file=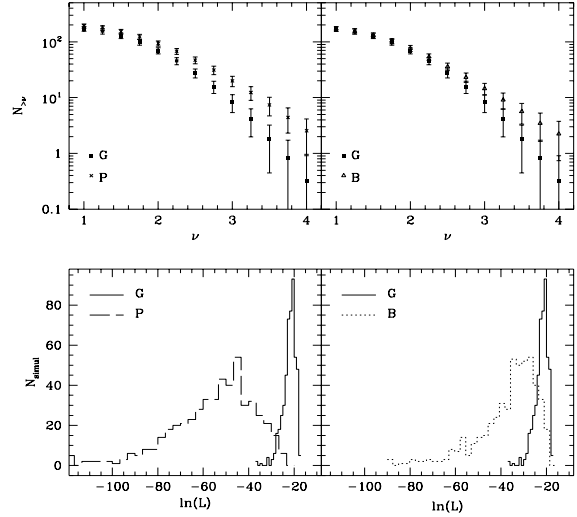, height=7.5cm, width=8cm}}
\caption{FIGURE 2. Top: $N_{>\nu}$ for thresholds $\nu$ in the range $1-4$.
Bottom: Distribution  of $Ln(Likelihood)$ corresponding to $N_{>\nu}$ for 
the Gaussian and non-Gaussian models (see text).
}
\end{figure}
%
}

\bsk
\baselineskip = 12pt
{\abstract \ni ACKNOWLEDGMENTS
This work has been supported by Spanish DGES Project PB95-1132-C02-02,
Spanish CICYT Acci\'on Especial ESP98-1545-E and USA-Spain Science 
and 
Technology Program, ref. 98138. JMD and RBB have been supported by a Spanish 
MEC fellowship.
}

\bsk
\baselineskip = 12pt


{\references \ni REFERENCES
\ssk

\ref Barreiro, R.B., Sanz, J.L., Mart\'\i nez-Gonz\'alez, E., Cay\'on, L.,
Silk, J., 1997, ApJ, 478, 1.

\ref Barreiro, R.B., Sanz, J.L., Mart\'\i nez-Gonz\'alez, E., Silk, J., 1998,
MNRAS, 296, 693.

\ref Cole, S. \& Weinberg, D.H., 1992, MNRAS, 259, 652.

\ref Coles, P. \& Barrow, J.D., 1987, MNRAS, 228, 407.

\ref Diego, J.M., Mart\'\i nez-Gonz\'alez, E., Sanz, J.L., Mollerach, S. \&
Mart\'\i nez, V.J., 1999, MNRAS, in press (astro-ph/9902134)

\ref Ferreira, P.G., Magueijo, J. \& G\'orski, K., 1998, ApJ, 503, L1.

\ref Gott III, J.R. et al., 1990, ApJ, 352, 1.


\ref Heavens, A.F., 1998, astro-ph/9804222.

\ref Hobson, M.P., Jones, A.W., Lasenby, A.N. \& Bouchet, F.R., 1998a, MNRAS,
300, 1.

\ref Hobson, M.P. et al., 1999a, MNRAS, in press (astro-ph/9810241)


\ref Hobson, M.P., Jones, A.W. \& Lasenby, A.N., 1999b, MNRAS, in press. 
(astro-ph/9810200)

\ref Kogut, A. et al., 1996, 464, L29.


\ref Luo, X., 1994, ApJ, 427, 71L.

\ref Pando, J., Valls-Gabaud, D., Fang, L., 1998, PRL, 81, 4568.

\ref Pompilio, M.P., Bouchet, F.R., Murante, G. \& Provenzale, A., 1995,
ApJ, 449, 1


\ref Tegmark, M. \& Efstathiou, G., 1996, MNRAS, 281, 1297.


}                      

\end{document}